# Flattened photon beams, an obsolete feature in modern linear accelerators

## E.I. Parsai[1*], E. Salari[1], D. Shvydka[1], J. Wan[2]

[1]Department of Radiation Oncology, University of Toledo Medical Center, 1325 Conference Drive, Toledo, OH 43614, USA
[2]Advanced Radiation Physics Service, Inc. Toledo, Ohio Metropolitan Area, Toledo, OH 43614, USA

## ABSTRACT

**Background:** With the advent of Intensity Modulated Radiotherapy (IMRT) and recently, Volumetric Modulated Arc Therapy (VMAT), treatment planning using Flattening Filter Free (FFF) beams can meet all of the energy requirements in radiation therapy clinics. Manufacturers of linear accelerators no longer need to install a flattening filter (FF) in gantry head. This study aims to provide evidence of the superiority of FFF to FF through both dosimetric measurements and clinical treatment plans. **Materials and Methods:** A 50×50×50cm$^3$ water phantom was created in the RayStation treatment planning system (TPS) for dosimetry comparisons. Flat beam profiles were generated using FFF beam through an optimization process for 10×10 to 30×30cm$^2$ field sizes. Next, a comparison of treatment plans was made using 21 Head and Neck and 14 Lung/Mediastinum treatment sites using 6MV and 6MV-FFF beams. **Results:** Using FFF beams, profiles with flatness and symmetry identical to or better than those of the flattened beams were produced. At the very edge of the optimized plans for FFF beams, horns had the highest gamma index deviation <1.5% of the normalized dose. For clinical plans evaluated, most of the mean doses to organs-at-risk (OAR) volumes receiving 5% to 30% of the prescription dose were reduced with FFF beams. **Conclusion:** These results indicate the feasibility of delivering flat beams with FFF quality and producing treatment plans with equal or higher qualities in PTV coverage while achieving better sparing of OAR which will allow escalation of target dose if desired. Plus, removing FF will simplify the gantry head and reduces quality assurance and machine maintenance efforts.

## INTRODUCTION

A flattening filter (FF) is designed to produce uniform dose distribution at a certain depth in a homogenous phantom, usually water. However, having a flat beam is not desirable for complex treatment plans. Therefore, beam-modifying devices such as compensators, wedges, and dynamic multileaf collimators (MLC) are used to shape beams. Over the past two decades, modern linear accelerators have been equipped with a flattening filter-free (FFF) feature, and a wealth of literature has demonstrated the advantages of FFF beams. Those, aside from dosimetric advantages which are the subject of this manuscript include but are not limited to its ability to produce treatment plans with sharper dose fall-off resulting in lower dose to normal structures in the vicinity of a target volume, and decreased radiation from head scatter and outside the treatment field since FF is identified as the most significant source of scatter radiation in gantry head [1-6]. Moreover, removing FF from the beam path results in a higher dose rate (1400 MU/cGy and 2400 MU/cGy for 6 MV FFF and 10 MV FFF beams respectively), leading to a shorter delivery time [7]. This decreased beam-on time is especially important for patients receiving Stereotactic Body Radiotherapy (SBRT) with gating, resulting in acceptable acute toxicity profiles and promising local control [7-9]. These advantages have been employed for numerous sites, including lung, liver, and brain [10-13] treatments. As a result, the FFF beams are widely used in SBRT and also in stereotactic radiotherapy (SRS) techniques where a smaller number of fractions with a higher dose per fraction is prescribed. In one study [2], the use of 6 MV-FFF beams was compared to 6 MV in plans produced for SRS treatments with improved conformity and better sparing of nearby critical structures, while reducing the beam-on time by roughly 43%. Furthermore, the removal of the FF helps establish much simpler configurations in the gantry of linac, which eliminates quality assurances to the filter and reduces expenses on building (from manufacturers' point of view) and purchasing (from clinical consumers' point of view) the machine.

A feature of the non-flat beam is that it presents highest intensity at the beam center in contrast to FF beam where typically a higher intensity is observed near the edges of the field known as horns. Using the MLCs through the sliding window technique in treatment planning software package, one has the ability to shape the beam fluency distribution across the field and deliver a desired dose distribution[5]. The majority of modern linear accelerators manufactured at the time of the writing, however, still provide flattened beams in addition to FFF photon beams. This study aims to provide the evidence that through inverse planning with VMAT delivery, no longer a flattened beam is needed, and the FF should be completely removed from the LINAC's head, thus reducing its complexity and to some degree the cost of manufacturing.

## MATERIALS AND METHODS

### Using non-flat photon beams to deliver a flat beam

For this study, Edge and TrueBeam linacs (Varian Medical Systems, Palo Alto, CA) with 6 MV-FFF, 6 MV, 10 MV-FFF, and 10 MV beams were utilized. Energies used were 10 MV flattened and 10 MV FFF from the TrueBeam and 6 MV flattened and 6MV FFF from the Edge linac. TrueBeam linac is equipped with a conventional 120 leaf MLC (60 pairs) with the central 20 cm having 5mm leaf width and the outer 20 cm having 10 mm leaf width with the maximum leaf speed of 2.5 cm/s. The Edge linac on the other hand is equipped with 120 HD MLC leaves with the central 8 cm having 2.5 mm leaf width, and the outer 14 cm with 5 mm leaf width providing a maximum IMRT field size of 32 cm × 22 cm. Flat beam profiles were generated for the 6MV-FFF energy using inverse planning with the sliding window technique and compared with profiles from 6MV beam. For this purpose, a 50×50×50 cm³ water phantom was created in the RayStation (Ver.8) (RaySearch Medical Laboratories AB, Stockholm, Sweden) treatment planning system (TPS) [14, 15]. Then beams with open square field sizes of 10×10, 20×20, and 30×30 cm² were defined by jaws and MLCs tracking the jaws at 100 cm SAD for both linacs at gantry angle 0°. The main optimization criterion for inverse plans was that of uniform dose to a plane with a thickness of 0.1 cm and areas equal to corresponding field sizes at 10 cm depth from the surface of the water.

For normalization purposes, the center of each plane was prescribed to receive 1 Gy. The optimization parameter "uniform dose" was utilized to guide the TPS to achieve the set goals by the MLCs sliding movement within the fields. After successfully producing uniform dose distribution on the plane, the "line dose" tool in RayStation TPS was used to get crossline and inline profiles [16, 17]. To obtain the beam profile, a line can be drawn across any of the regions of interest by this tool. In this case, profiles for different field sizes across the central axis and vertical to the sagittal plane of the water phantom at 10 cm depth from the surface of the water were gathered for data analysis. Then flatness was calculated based on equation 1:

$$Flatness = \frac{(D_{max} - D_{min})}{(D_{max} + D_{min})} \times 100 \qquad (1)$$

Where $D_{max}$ and $D_{min}$ are the maximum and minimum doses along with the profile within the central 80% of the field.

The gamma index was also calculated based on 3%/3mm objectives by using an in-house developed code written in Python3 (Python Software Foundation) to compare the profiles generated by non-flat beams with those generated using flat beams.

### Clinical treatment plans comparison (6 MV vs 6 MV FFF)

This comparison aims to verify the feasibility of creating identical or even higher-quality plans with FFF beams. For this purpose, 21 Head and Neck (H&N) patients and 14 Lung/Mediastinum patients who were previously treated with 6 MV photon beams were selected. New plans with 6 MV FFF photon beams were generated for comparison. All new completed plans with 6 MV FFF beams achieved a similar percentage coverage of at least one planning target volume (PTV) level. Ethical approval was obtained for this research from the Internal Review Board (IRB) of the University of Toledo (UT-300579) on April 2nd, 2020.

The simultaneous integrated boost technique was used for both Lung/Mediastinum and H&N treatment plans. Lung/Mediastinum plans have one to three PTVs with different dose levels (30 Gy to 60 Gy) delivered in 10 fractions. For the H&N cases with a total of three targets, prescription doses of 54 to 66 Gy in 30 fractions were used, or plans were only designed for one target with a prescription of 36 Gy or 40 Gy in 10 fractions. Depending on the size of the target, 2 or 4 arcs were used for both H&N and Lung/Mediastinum plans. Most objectives and constraints used for plan optimization remained unchanged, only a few extra objectives were defined to meet the demand for the equivalent coverage of PTVs. Average differences between plans with non-flat beams and with flat beams for maximum doses, mean doses, and volumes receiving 5%, 10%, 20%, and 30% of the prescription dose for organs-at-risk (OAR) were selected to evaluate the results.

The objective is choosing the low-dose level irradiation to OAR for an investigation came from the knowledge that the greatest advantage of the non-flat beam against the conventional flat beam would be a fast dose fall-off beyond the target. Consequently, it is expected that less contribution of dose to normal tissues should be observed in the results. RadCalc™ (Ver6.4) (LifeLine Software, Inc., LAP Group) was used as an independent monitor unit verification calculation to confirm the accuracy of dose calculations in the TPS.

# RESULTS

### Using non-flat photon beams to deliver a flat beam

Crossline and inline profiles for both FFF beams, and flat beams overlaid on top of each other are shown in figures 1 to 5. Each line profile was extracted from the RayStation TPS in Microsoft Excel (Ver. 2016) format datasheet, which will allow obtaining point dose values along the line. The gamma index line is shown in each graph of figures 1 to 5 and is multiplied by 20 for clarity.

Equation 1 was utilized to calculate the flatness of all profiles, with results presented in table 1. Due to jaw opening limits on the Edge machine, 30×30 cm² fields were not generated for both 10 MV FFF and 10MV beams.

The results from the initial part of this research already indicated that it is highly feasible to deliver a flat beam with a non-flat beam.

Also, the dose distribution of 6MV FFF and 6MV beams as illustrated in figures 6, 7 & 8 indicates a sharp dose fall of FFF beams beyond the target(s).

All doses calculated for both 6MV FFF and 6MV with RadCalc were within ±2% of the doses calculated by TPS. Moreover, for each treatment site, the average delivery time of 6 MV FFF was compared with the average delivery time results of 6MV shown in figure 9. The maximum dose rate was used to achieve the fastest delivery time for each plan in TPS. As shown in figure 10, more monitor units were needed to generate a uniform dose distribution in the PTV region using non-flat beams; however, this did not lead to longer treatment times for beam with FFF energies.

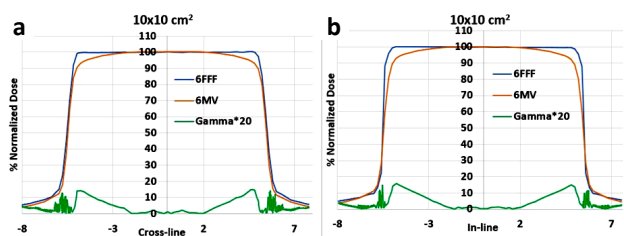

**Figure 1.** Field size 10 × 10 cm² for 6 MV FFF and 6 MV beam at 10 cm depth and SAD 100cm. **a:** Cross line, **b:** Inline.

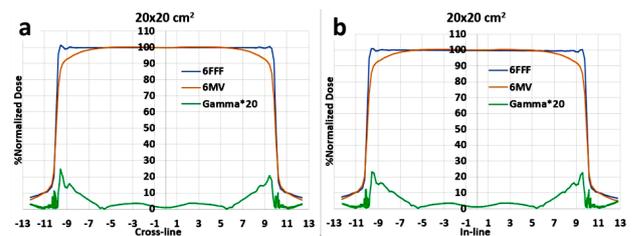

**Figure 2.** Field size 20 × 20 cm² for 6 MV FFF and 6 MV beam at 10 cm depth and SAD 100cm. **a:** Cross line, **b:** Inline.

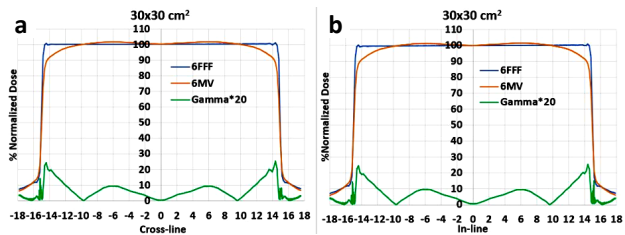

**Figure 3.** Field size 30 × 30 cm² for 6 MV FFF and 6 MV beam at 10 cm depth and SAD 100cm. **a:** Cross line, **b:** Inline.

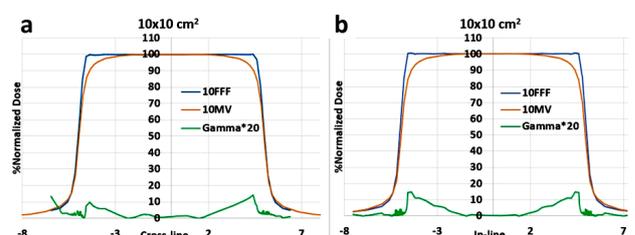

**Figure 4.** Field size 10 × 10 cm² for 10 MV FFF and 10 MV beam at 10 cm depth and SAD 100cm. **a:** Cross line, **b:** Inline.

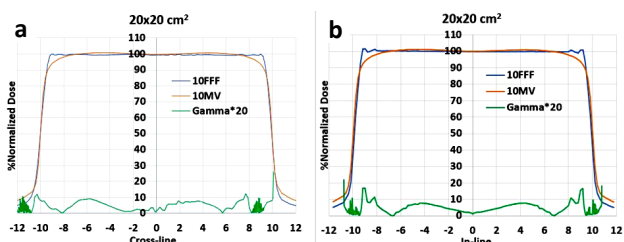

**Figure 5.** Field size 20 × 20 cm² for 10 MV FFF and 10 MV beam at 10 cm depth and SAD 100cm. **a:** Cross line, **b:** Inline.

**Table 1.** Flatness of 6MV vs 6MV FFF beams for field sizes of 10×10 cm², 20×20cm², 30×30 cm² and 10MV vs 10MV FFF for field sizes of 10×10 cm² and 20×20 cm². Abbreviation of FFF refers to Flattening Filter Free beams.

| FS/Energy | Crossline | | | | Inline | | | |
|---|---|---|---|---|---|---|---|---|
| | 6MV FFF | 6MV | 10MV FFF | 10MV | 6MV FFF | 6MV | 10MV FFF | 10MV |
| 10x10 cm2 | 0.264 | 1.945 | 0.237 | 2.168 | 0.244 | 2.069 | 0.163 | 2.602 |
| 20x20 cm2 | 0.103 | 2.038 | 0.412 | 1.589 | 0.415 | 2.336 | 0.323 | 1.733 |
| 30x30 cm2 | 0.135 | 2.066 | N/A | N/A | 0.303 | 2.276 | N/A | N/A |

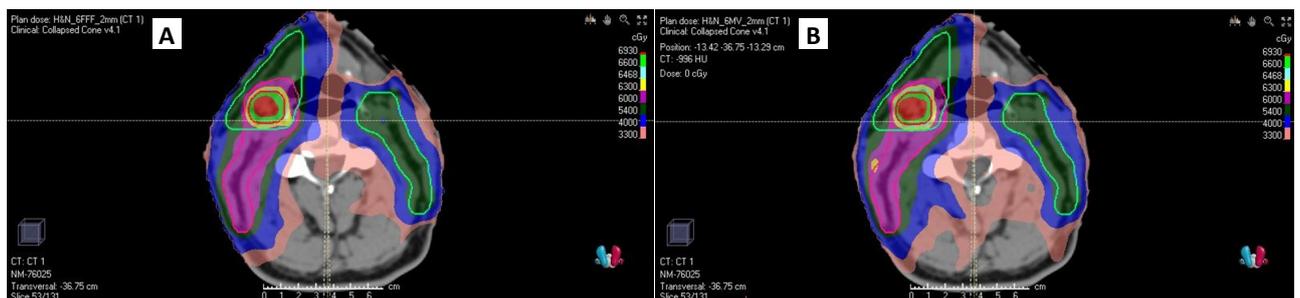

**Figure 6.** Example of the Axial-view dose distribution in Head and Neck cancer treatment plans. **A)** 6MV FFF, and **B)** 6MV- PTV Dose levels of 66, 60 and 54 Gy.

**Table 2.** Differences in average maximum doses (Gy), mean doses (Gy), and volumes receiving 5%, 10%, 20%, and 30% of the prescription dose between 6 MV and 6 MV FFF for each organ at risk. **a)** Head & Neck cases; **b)** Lung/Mediastinum cases. The abbreviation of FFF refers to Flattening Filter Free.

| a) | Brainstem | Spinal Cord | Esophagus | Larynx | Left Parotid | Right Parotid | Trachea |
|---|---|---|---|---|---|---|---|
| V5% (cc) | 0.00 | 0.22 | 0.00 | -0.13 | 0.00 | 0.00 | 0.00 |
| V10% (cc) | -0.14 | 0.00 | 0.00 | 0.00 | -0.46 | 0.81 | 0.00 |
| V20% (cc) | -0.29 | -0.24 | 0.00 | 0.00 | -0.23 | 0.00 | 0.00 |
| V30% (cc) | -0.16 | -1.09 | -0.17 | -0.16 | -0.71 | 0.26 | -0.20 |
| Mean Dose (Gy) | 0.00 | -0.34 | -0.36 | -0.55 | 2.00 | 2.25 | -16.50 |
| Max Dose (Gy) | -35.24 | -39.71 | -13.33 | -21.70 | -3.71 | -3.44 | 7.75 |
| b) | Spinal Cord | Esophagus | Heart | Lungs | Trachea & Carina | | |
| V5% (cc) | -0.94 | -0.53 | -6.65 | -33.67 | -0.77 | | |
| V10% (cc) | -0.18 | 0.42 | -8.00 | -33.53 | -1.45 | | |
| V20% (cc) | -0.68 | -0.50 | -7.58 | -32.52 | -1.55 | | |
| V30% (cc) | -1.84 | -0.86 | -9.27 | -30.72 | -1.43 | | |
| Mean Dose (Gy) | -35.22 | -34.36 | -24.00 | -11.43 | -74.36 | | |
| Max Dose (Gy) | -19.71 | -38.14 | -216.21 | -8.00 | 6.93 | | |

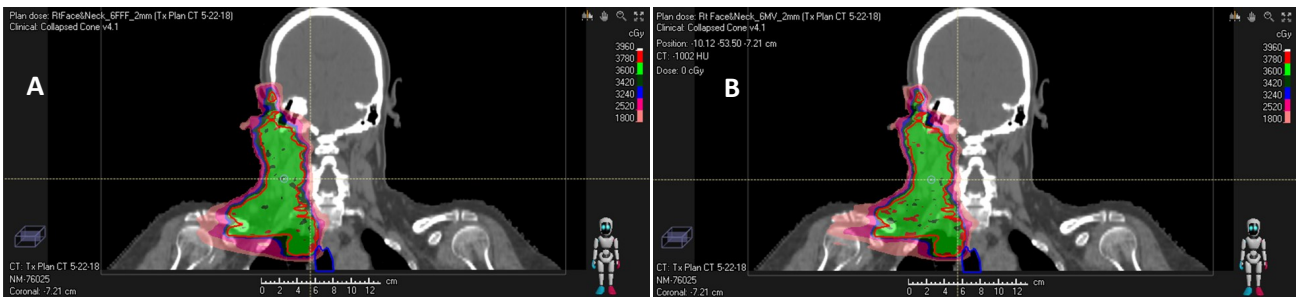

**Figure 7.** Example of the Coronal-view dose distribution in Head and Neck cancer treatment plans. A) 6MV FFF, and B) 6MV- Dose level 36 Gy.

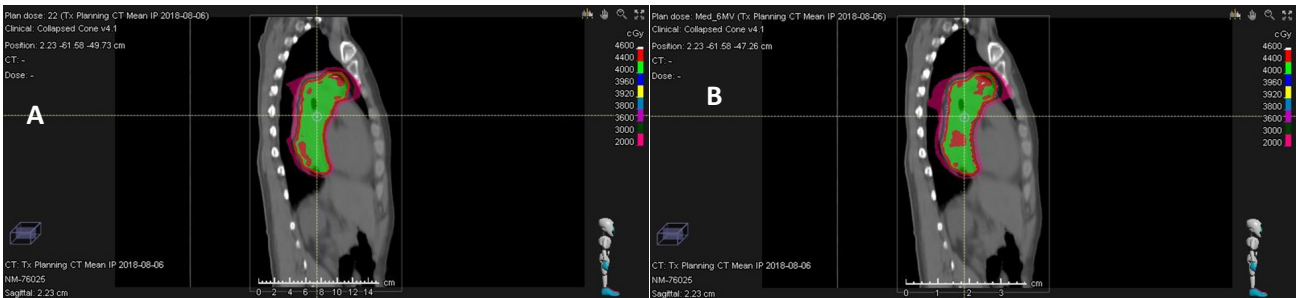

**Figure 8.** Example of the sagittal-view dose distribution in a Lung/Mediastinum cancer treatment plan. **A)** 6MV FFF and **B)** 6MV- Dose level 40Gy.

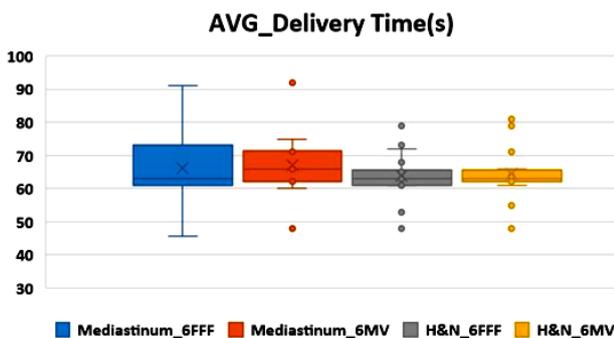

**Figure 9.** Average delivery time(s) for Lung/Mediastinum and H&N treatment plans while the maximum available dose rate on the Linac was utilized. Whisker chart shows the distribution of data into quartiles, the line and X within the box show the median and mean values respectively. Dots outside the box are outliers. Y-axis is the average delivery time (second).

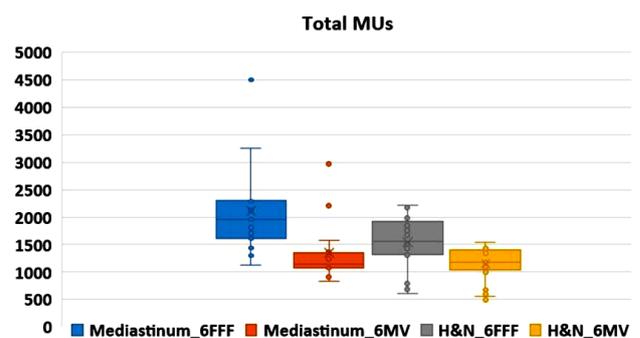

**Figure 10.** Total number of MUs for Lung/Mediastinum and H&N treatment plans while the maximum available dose rate on the Linac was utilized. Whisker chart shows the distribution of data into quartiles, the line and X within the box show the median and mean values respectively. Dots outside the box are outliers. Y-axis is the Total MU.

## DISCUSSION

Using a commissioned RayStation TPS on Varian linacs, the FFF beams were optimized through a sliding window inverse planning process to confirm the ability to deliver a flat beam with various field sizes. Large field sizes (10×10, 20×20, and 30×30 cm$^2$) were chosen for this study to prove the concept that a TPS can generate flat profiles with FFF beams, thus mimicking the effect of a flattened beam. As shown in table 1, for beam profiles computed through the TPS, the flatness of dose was superior for both FFF beams in both inline and cross line in contrast to the flat beams. As evident from figures 1 to 5, minor deviations in the dose profiles were observed at the field edges, where instead of a gradual decrease of the flattened beam profiles the optimized FFF beam plans resulted in sharper edge drop-offs and slight "horn" features. The highest dose deviation in those regions never exceeded 5% of the normalized dose. Gamma passing rates are also shown in figures 1 to 5 (note the multiplication factor of 20 used for clarity) and point also to excellent agreement between the profiles except for a few points coinciding with the "horn" locations. Small percentage differences in flatness between flat beams and FFF within a specific square plane demonstrate the feasibility of using a non-flat beam to generate the flat dose distribution with the sliding window technique. Our result is in good agreement with Potter *et al.* findings which demonstrated producing a modulated flat beam using a FFF beam is practicable [4].

A set of H&N and Lung/Mediastinum plans were used to illustrate the capabilities of FFF beams in achieving both the superior dose conformity to the target and faster dose fall-off outside the target volumes. For most of the H&N and all Lung/Mediastinum plans volumes of each OAR adjacent to targets receiving low doses in FFF beams-based plans were reduced, as shown in table 2. Also, from figures 6, 7 & 8, it is obvious that the FFF beams achieved uniformity within the region of the target(s) as good or better than conventional flat beams.

Furthermore, mean doses for OAR decreased, maximum doses increased slightly in the high dose level target, and the maximum dose of OAR declined when FFF beams were utilized. Mean doses of both sides of the parotids slightly increased for non-flat beams since some parts of these organs were in the PTV region. Similar trends with much more significant dose reduction in OAR were found for Lung/Mediastinum treatment plans. The only observed exception was in the trachea, which was a part of PTVs for one patient, as some hot spots were included in those areas, resulting in a higher max dose. Several studies were conducted to compare FFF vs FF beams for different treatment sites [10, 18, 11-13]. Our study had similar outcomes which are in parallel with other findings.

Figures 6 and 7 show the dose distribution for one example of each group of treated sites.

These results indicate that it is feasible to deliver a flat beam with a FFF quality and produce treatment plans with escalated total doses while sparing OAR. Albeit non-flat beams might generate higher maximum doses (hot spots) in the whole plan, an increase of less than 3% of the maximum dose should not cause any additional biological complications. Trading a very small escalation of a maximum point dose with preventing OAR from receiving an extra low dose seems a good compromise. Some increase in delivered MU's for FFF plans is another trade-off in achieving higher quality complex plans as shown in figure 10. Salari *et al.* [19] also have recently shown that the decrease in off-axis ratio, a characteristic of FFF beams results in MU increase to generate the same uniform PTV coverage which is also in good agreement with the Cashmore's result [20]. Similarly, we can conclude that rapid dose fall-off in FFF beams generally require more MUs to produce the same PTV coverage as flat beams.

As shown in figure 9, unlike other TPS [1, 4, 21], no significant difference was observed in the delivery time of flat vs non-flat beams. This is due to RayStation's optimization algorithm which is capable of providing similar delivery times for both flat and non-flat beams by adjusting other variables such as dose rate and gantry speed to deliver a specific amount of MU in the VMAT technique. It was also shown that the delivery time completely hinges on the ability of the optimization algorithm in VMAT technique where gantry speed and dose rate are two more variables compared to the IMRT technique.

## CONCLUSIONS

Using a commissioned RayStation TPS on the Varian linacs, the FFF beams were optimized through sliding window inverse planning process to confirm its capability to deliver flat beams with various field sizes from 10×10 to 30×30 cm$^2$. The study also demonstrated the superiority of FFF beams to flat beams by comparison of the dosimetric characteristics of beam sets, and also by comparing clinical treatment plans. With identical coverages of PTVs, lower doses to OARs were achieved with FFF beams in plans presented for H&N and mediastinum. As a result, the complete removal of the flattening filter from the gantry head of modern linear accelerators is possible and recommended as it eliminates additional quality assurances for filtered beams while lowering the added complexities in electronics and expenses at the time of manufacturing.

**ACKNOWLEDGMENT**
*None.*

***Conflict of Interests:*** None to report.

**Ethical consideration**: Ethical approval for the research was obtained from the University of Toledo (UT-300579).

**Funding**: None to report.

**Author contribution**: All authors designed, conceived, analysis the collected data. They contributed and performed the needful in writing the paper. *E.I. Parsai*: Conceived and designed the analysis, collected the data, contributed data, performed the analysis, and wrote the paper. *E. Salari*: Conceived and designed the analysis, collected the data, contributed data, performed the analysis, wrote the paper. *D. Shvydka*: Conceived and designed the analysis, wrote the paper. J. Wan: Conceived and designed the analysis

## REFERENCES

1. Fu W, Dai J, Hu Y, Han D, Song Y (2004) Delivery time comparison for intensity-modulated radiation therapy with/without flattening filter: a planning study. *Phys Med Biol*, **49(8)**: 1535-1547.
2. Lai Y, Chen S, Xu C, Shi L, Fu L, Ha H, *et al.* (2017) Dosimetric superiority of flattening filter free beams for single-fraction stereotactic radiosurgery in single brain metastasis. *Oncotarget*, **8(21)**: 35272-35279.
3. Pichandi A, Ganesh KM, Jerin A, Balaji K, Kilara G (2014) Analysis of physical parameters and determination of inflection point for Flattening Filter Free beams in medical linear accelerator. *Rep Pract Oncol Radiother*, **19(5)**:322-331.
4. Potter NJ, Lebron S, Li JG, Liu C, Lu B (2019) Feasibility study of using flattening-filter-free photon beams to deliver conventional flat beams. *Med Dosim*, **44(4)**: e25-e31.
5. Titt U, Vassiliev ON, Ponisch F, Dong L, Liu H, Mohan R (2006) A flattening filter free photon treatment concept evaluation with Monte Carlo. *Med Phys*, **33(6)**: 1595-1602.
6. Titt U, Vassiliev ON, Ponisch F, Kry SF, Mohan R (2006) Monte Carlo study of backscatter in a flattening filter free clinical accelerator. *Med Phys*, **33(9)**: 3270-3273.
7. Prendergast BM, Fiveash JB, Popple RA, Clark GM, Thomas EM, Minnich DJ, *et al.* (2013) Flattening filter-free linac improves treatment delivery efficiency in stereotactic body radiation therapy. *Journal of Applied Clinical Medical Physics*, **14(3)**: 4126-4126.
8. Purdie TG, Bissonnette JP, Franks K, Bezjak A, Payne D, Sie F, *et al.* (2007) Cone-beam computed tomography for on-line image guidance of lung stereotactic radiotherapy: localization, verification, and intrafraction tumor position. *Int J Radiat Oncol Biol Phys*, **68(1)**: 243-252.
9. Scorsetti M, Alongi F, Castiglioni S, Clivio A, Fogliata A, Lobefalo F, *et al.* (2011) Feasibility and early clinical assessment of flattening filter free (FFF) based stereotactic body radiotherapy (SBRT) treatments. *Radiat Oncol*, **6**: 113.
10. Arslan A and Sengul B (2020) Comparison of radiotherapy techniques with flattening filter and flattening filter-free in lung radiotherapy according to the treatment volume size. *Scientific Reports*, **10(1)**: 8983.
11. Nakano H, Minami K, Yagi M, Imaizumi H, Otani Y, Inoue S, *et al.* (2018) Radiobiological effects of flattening filter–free photon beams on A549 non-small-cell lung cancer cells. *Journal of Radiation Research*, **59(4)**: 442-445.
12. Parsai E, Pearson D, Kvale T (2007) Consequences of removing the flattening filter from linear accelerators in generating high dose rate photon beams for clinical applications: A Monte Carlo study verified by measurement. *Nuclear Instruments and Methods in Physics Research Section B: Beam Interactions with Materials and Atoms*, **261**: 755-759.
13. Rieber J, Tonndorf-Martini E, Schramm O, Rhein B, Stefanowicz S, Kappes J, *et al.* (2016) Radiosurgery with flattening-filter-free techniques in the treatment of brain metastases : Plan comparison and early clinical evaluation. *Strahlenther Onkol*, **192(11)**:789-796.
14. RayStation 8A RayPhysics Manual (2018) RaySearch Laboratories.
15. RayStation 8A User Manual (2018) RaySearch Laboratories.
16. Beam Commissioning Data Specification RayStation 8A (2018) RaySearch Laboratories.
17. RayStation 8A Reference Manual (2018) RaySearch Laboratories.
18. Kim G, Rice R, Lawson J, Murphy K, Pawlicki T (2012) Stereotactic Radiosurgery With FFF Mode Photon Beams. *Int J Radiat Oncol Biol Phys*, **84(3)**: S823.
19. Salari E, Parsai EI, Shvydka D, Sperling NN (2022) Evaluation of parameters affecting gamma passing rate in patient-specific QAs for multiple brain lesions IMRS treatments using ray-station treatment planning system. *J Appl Clin Med Phys*, **23(1)**: e13467.
20. Cashmore J (2008) The characterization of unflattened photon beams from a 6 MV linear accelerator. *Phys Med Biol*, **53(7)**:1933-1946.
21. Thomas EM, Popple RA, Prendergast BM, Clark GM Dobelbower MC, Fiveash JB (2013) Effects of flattening filter-free and volumetric-modulated arc therapy delivery on treatment efficiency. *J Appl Clin Med Phys*, [Internet]. [PMC5714642]; **14(6)**: 4328.